\documentclass[12pt,a4paper]{article}
\pdfoutput=1
\usepackage{ifthen} 
\newboolean{pdflatex}
\setboolean{pdflatex}{true} 

\newboolean{articletitles}
\setboolean{articletitles}{true} 

\newboolean{uprightparticles}
\setboolean{uprightparticles}{false} 

\newboolean{inbibliography}
\setboolean{inbibliography}{false} 

\usepackage[top=1in, bottom=1.25in, left=1in, right=1in]{geometry}

\usepackage{microtype}
\usepackage{lineno}  
\usepackage{xspace} 
                    
\usepackage{caption}

\usepackage{graphicx}  
\usepackage{color}
\usepackage{colortbl}
\graphicspath{{./figs/}} 

\usepackage{amsmath} 
\usepackage{amssymb}
\usepackage{amsfonts}
\usepackage{upgreek} 

\newcommand*\patchAmsMathEnvironmentForLineno[1]{%
\expandafter\let\csname old#1\expandafter\endcsname\csname #1\endcsname
\expandafter\let\csname oldend#1\expandafter\endcsname\csname
end#1\endcsname
 \renewenvironment{#1}%
   {\linenomath\csname old#1\endcsname}%
   {\csname oldend#1\endcsname\endlinenomath}%
}
\newcommand*\patchBothAmsMathEnvironmentsForLineno[1]{%
  \patchAmsMathEnvironmentForLineno{#1}%
  \patchAmsMathEnvironmentForLineno{#1*}%
}
\AtBeginDocument{%
\patchBothAmsMathEnvironmentsForLineno{equation}%
\patchBothAmsMathEnvironmentsForLineno{align}%
\patchBothAmsMathEnvironmentsForLineno{flalign}%
\patchBothAmsMathEnvironmentsForLineno{alignat}%
\patchBothAmsMathEnvironmentsForLineno{gather}%
\patchBothAmsMathEnvironmentsForLineno{multline}%
\patchBothAmsMathEnvironmentsForLineno{eqnarray}%
}

\usepackage{hyperref}    
\usepackage[all]{hypcap} 

\usepackage{xspace} 
\usepackage{upgreek}

\def\lhcb {\mbox{LHCb}\xspace}

\def\babar  {\mbox{BaBar}\xspace}
\def\belle  {\mbox{Belle}\xspace}

\def\dzero  {\mbox{D0}\xspace}

\def\lhc    {\mbox{LHC}\xspace}

\def\MagUp {\mbox{\em Mag\kern -0.05em Up}\xspace}

\ifthenelse{\boolean{uprightparticles}}
{

 \def\Pmu         {\ensuremath{\upmu}\xspace}

 \def\Ppi         {\ensuremath{\uppi}\xspace}

 \def\Ppsi        {\ensuremath{\uppsi}\xspace}

 \def\PDelta      {\ensuremath{\Delta}\xspace}                 
 \def\PXi      {\ensuremath{\Xi}\xspace}                 
 \def\PLambda      {\ensuremath{\Lambda}\xspace}                 
 \def\PSigma      {\ensuremath{\Sigma}\xspace}                 
 \def\POmega      {\ensuremath{\Omega}\xspace}                 
 \def\PUpsilon      {\ensuremath{\Upsilon}\xspace}

 \def\PB      {\ensuremath{\mathrm{B}}\xspace}                 
                  
 \def\PD      {\ensuremath{\mathrm{D}}\xspace}

 \def\PJ      {\ensuremath{\mathrm{J}}\xspace}                 
 \def\PK      {\ensuremath{\mathrm{K}}\xspace}

 \def\Pb      {\ensuremath{\mathrm{b}}\xspace}                 
 \def\Pc      {\ensuremath{\mathrm{c}}\xspace}

 \def\Pi      {\ensuremath{\mathrm{i}}\xspace}

}
{

 \def\Pmu         {\ensuremath{\mu}\xspace}

 \def\Ppi         {\ensuremath{\pi}\xspace}

 \def\Ppsi        {\ensuremath{\psi}\xspace}                 
                  
 \mathchardef\PDelta="7101
 \mathchardef\PXi="7104
 \mathchardef\PLambda="7103
 \mathchardef\PSigma="7106
 \mathchardef\POmega="710A
 \mathchardef\PUpsilon="7107
                  
 \def\PB      {\ensuremath{B}\xspace}                 
                  
 \def\PD      {\ensuremath{D}\xspace}

 \def\PJ      {\ensuremath{J}\xspace}                 
 \def\PK      {\ensuremath{K}\xspace}

 \def\Pb      {\ensuremath{b}\xspace}                 
 \def\Pc      {\ensuremath{c}\xspace}

 \def\Pi      {\ensuremath{i}\xspace}

}

\makeatletter
\ifcase \@ptsize \relax
  \newcommand{\miniscule}{\@setfontsize\miniscule{4}{5}}
\or
  \newcommand{\miniscule}{\@setfontsize\miniscule{5}{6}}
\or
  \newcommand{\miniscule}{\@setfontsize\miniscule{5}{6}}
\fi
\makeatother

\DeclareRobustCommand{\optbar}[1]{\shortstack{{\miniscule (\rule[.5ex]{1.25em}{.18mm})}
  \\ [-.7ex] $#1$}}

\def\mup        {{\ensuremath{\Pmu^+}}\xspace}
\def\mun        {{\ensuremath{\Pmu^-}}\xspace} 
\def\mumu       {{\ensuremath{\Pmu^+\Pmu^-}}\xspace}

\def\cquark    {{\ensuremath{\Pc}}\xspace}

\def\bquark    {{\ensuremath{\Pb}}\xspace}

\def\pion   {{\ensuremath{\Ppi}}\xspace}

\def\pip    {{\ensuremath{\pion^+}}\xspace}
\def\pim    {{\ensuremath{\pion^-}}\xspace}
\def\pipm   {{\ensuremath{\pion^\pm}}\xspace}

\def\kaon    {{\ensuremath{\PK}}\xspace}
  \def\Kbar    {{\kern 0.2em\overline{\kern -0.2em \PK}{}}\xspace}

\def\KorKbar    {\kern 0.18em\optbar{\kern -0.18em K}{}\xspace}
\def\Kz      {{\ensuremath{\kaon^0}}\xspace}
\def\Kzb     {{\ensuremath{\Kbar{}^0}}\xspace}
\def\Kp      {{\ensuremath{\kaon^+}}\xspace}
\def\Km      {{\ensuremath{\kaon^-}}\xspace}

\def\KS      {{\ensuremath{\kaon^0_{\mathrm{ \scriptscriptstyle S}}}}\xspace}

  \def\Dbar    {{\kern 0.2em\overline{\kern -0.2em \PD}{}}\xspace}
\def\D       {{\ensuremath{\PD}}\xspace}

\def\DorDbar    {\kern 0.18em\optbar{\kern -0.18em D}{}\xspace}
\def\Dz      {{\ensuremath{\D^0}}\xspace}

\def\Dp      {{\ensuremath{\D^+}}\xspace}
\def\Dm      {{\ensuremath{\D^-}}\xspace}

\def\Dmp     {{\ensuremath{\D^\mp}}\xspace}

\def\Dstarp  {{\ensuremath{\D^{*+}}}\xspace}

\def\B       {{\ensuremath{\PB}}\xspace}
\def\Bbar    {{\ensuremath{\kern 0.18em\overline{\kern -0.18em \PB}{}}}\xspace}

\def\BorBbar    {\kern 0.18em\optbar{\kern -0.18em B}{}\xspace}

\def\Bu      {{\ensuremath{\B^+}}\xspace}
\def\Bub     {{\ensuremath{\B^-}}\xspace}
\def\Bp      {{\ensuremath{\Bu}}\xspace}
\def\Bm      {{\ensuremath{\Bub}}\xspace}
\def\Bpm     {{\ensuremath{\B^\pm}}\xspace}

\def\jpsi     {{\ensuremath{{\PJ\mskip -3mu/\mskip -2mu\Ppsi\mskip 2mu}}}\xspace}

  \def\Y#1S{\ensuremath{\PUpsilon{(#1S)}}\xspace}

\def\Lbar        {{\ensuremath{\kern 0.1em\overline{\kern -0.1em\PLambda}}}\xspace}
\def\LorLbar    {\kern 0.18em\optbar{\kern -0.18em \PLambda}{}\xspace}

\def\BF         {{\ensuremath{\mathcal{B}}}\xspace}

\def\BR         {\BF}

\def\to                 {\ensuremath{\rightarrow}\xspace}

\def\eps   {{\ensuremath{\varepsilon}}\xspace}

\def\CP                {{\ensuremath{C\!P}}\xspace}

\newcommand{\ACP}{{\ensuremath{{\mathcal{A}}^{\CP}}}\xspace}

\def\AT#1     {\ensuremath{A_{\mathrm{T}}^{#1}}\xspace}

\def\C#1      {\ensuremath{\mathcal{C}_{#1}}\xspace}                       
\def\Cp#1     {\ensuremath{\mathcal{C}_{#1}^{'}}\xspace}                   
\def\Ceff#1   {\ensuremath{\mathcal{C}_{#1}^{\mathrm{(eff)}}}\xspace}        
\def\Cpeff#1  {\ensuremath{\mathcal{C}_{#1}^{'\mathrm{(eff)}}}\xspace}     
\def\Ope#1    {\ensuremath{\mathcal{O}_{#1}}\xspace}                       
\def\Opep#1   {\ensuremath{\mathcal{O}_{#1}^{'}}\xspace}

\newcommand{\tev}{\ifthenelse{\boolean{inbibliography}}{\ensuremath{~T\kern -0.05em eV}\xspace}{\ensuremath{\mathrm{\,Te\kern -0.1em V}}}\xspace}
\newcommand{\gev}{\ensuremath{\mathrm{\,Ge\kern -0.1em V}}\xspace}
\newcommand{\mev}{\ensuremath{\mathrm{\,Me\kern -0.1em V}}\xspace}
\newcommand{\kev}{\ensuremath{\mathrm{\,ke\kern -0.1em V}}\xspace}
\newcommand{\ev}{\ensuremath{\mathrm{\,e\kern -0.1em V}}\xspace}
\newcommand{\gevc}{\ensuremath{{\mathrm{\,Ge\kern -0.1em V\!/}c}}\xspace}
\newcommand{\mevc}{\ensuremath{{\mathrm{\,Me\kern -0.1em V\!/}c}}\xspace}
\newcommand{\gevcc}{\ensuremath{{\mathrm{\,Ge\kern -0.1em V\!/}c^2}}\xspace}
\newcommand{\gevgevcccc}{\ensuremath{{\mathrm{\,Ge\kern -0.1em V^2\!/}c^4}}\xspace}
\newcommand{\mevcc}{\ensuremath{{\mathrm{\,Me\kern -0.1em V\!/}c^2}}\xspace}

\def\mum  {\ensuremath{{\,\upmu\mathrm{m}}}\xspace}

\def\invfb   {\ensuremath{\mbox{\,fb}^{-1}}\xspace}

\def\gsim{{~\raise.15em\hbox{$>$}\kern-.85em
          \lower.35em\hbox{$\sim$}~}\xspace}
\def\lsim{{~\raise.15em\hbox{$<$}\kern-.85em
          \lower.35em\hbox{$\sim$}~}\xspace}

\def\sPlot{\mbox{\em sPlot}\xspace}

\def\ptot       {\mbox{$p$}\xspace}
\def\pt         {\mbox{$p_{\mathrm{ T}}$}\xspace}

\def\evtgen     {\mbox{\textsc{EvtGen}}\xspace}

\def\geant      {\mbox{\textsc{Geant4}}\xspace}

\def\photos     {\mbox{\textsc{Photos}}\xspace}

\def\pythia     {\mbox{\textsc{Pythia}}\xspace}

\def\tell1  {TELL1\xspace}
\def\ukl1   {UKL1\xspace}

\usepackage{cite} 
\usepackage{mciteplus}

\usepackage{longtable} 

\begin{document}

\newcommand{\Bk}{\Bu\to\jpsi\Kp}
\newcommand{\Bpi}{\Bu\to\jpsi\pip}
\newcommand{\rpik}{\mathcal{R}_{\pi/K}}
\newcommand{\apik}{\Delta\ACP}
\newcommand{\api}{A_{\rm CP}^{\pi}}
\newcommand{\ak}{A_{\rm CP}^{K}}
\newcommand{\red}{\textcolor[rgb]{1,0,0}}

\renewcommand{\thefootnote}{\fnsymbol{footnote}}
\setcounter{footnote}{1}

\begin{titlepage}
\pagenumbering{roman}

\vspace*{-1.5cm}
\centerline{\large EUROPEAN ORGANIZATION FOR NUCLEAR RESEARCH (CERN)}
\vspace*{1.5cm}
\noindent
\begin{tabular*}{\linewidth}{lc@{\extracolsep{\fill}}r@{\extracolsep{0pt}}}
\ifthenelse{\boolean{pdflatex}}
{\vspace*{-2.7cm}\mbox{\!\!\!\includegraphics[width=.14\textwidth]{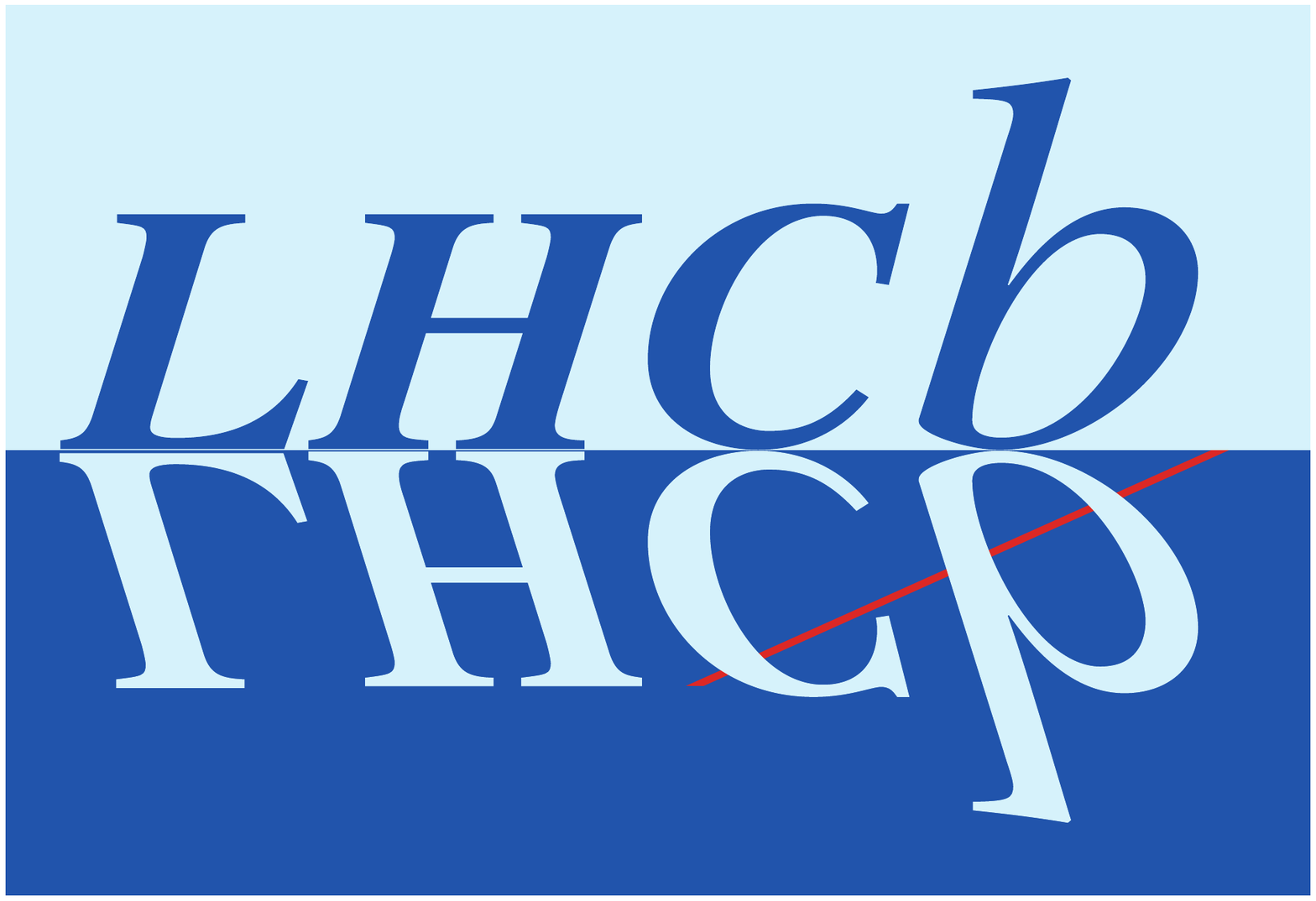}} & &}
{\vspace*{-1.2cm}\mbox{\!\!\!\includegraphics[width=.12\textwidth]{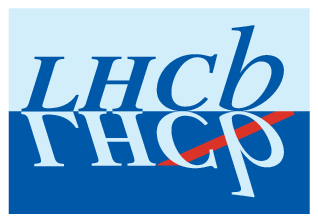}} & &}
\\
 & & CERN-EP-2016-298 \\  
 & & LHCb-PAPER-2016-051 \\  
 & & March 13, 2017 \\

\end{tabular*}

\vspace*{3.0cm}

{\normalfont\bfseries\boldmath\huge
\begin{center}
 Measurement of the ratio of branching fractions and 
 difference in \CP asymmetries of the decays $\Bpi$ and $\Bk$
\end{center}
}

\vspace*{2.0cm}

\begin{center}
The LHCb collaboration\footnote{Authors are listed at the end of this paper.}
\end{center}

\vspace{\fill}

\begin{abstract}
  \noindent
The ratio of branching fractions and the difference in \CP asymmetries of the decays 
$\Bpi$ and $\Bk$ are measured using a data sample of $pp$ collisions collected
by the \lhcb experiment, corresponding to an integrated luminosity of $3 \invfb$
at centre-of-mass energies of 7 and $8\tev$. The results are
\begin{equation*}
{\BR(\Bpi)\over\BR(\Bk)}=(3.83\pm0.03\pm0.03 ) \times10^{-2}
\,,
\end{equation*}
\begin{equation*}
\ACP(\Bpi)-\ACP(\Bk)=(1.82\pm0.86\pm0.14)\times 10^{-2}
, 
\end{equation*}
where the first uncertainties are statistical and the second are systematic.
Combining this result with a recent \lhcb measurement of $\ACP(\Bk)$ provides
the most precise estimate to date of \CP violation in the decay $\Bpi$,
\begin{equation*}
\ACP(\Bpi) = (1.91\pm0.89\pm0.16)\times10^{-2}
.
\end{equation*}
  
\end{abstract}

\vspace*{1.0cm}

\begin{center} Published in JHEP 03 (2017) 036. 
\end{center}

\vspace{\fill}

{\footnotesize \centerline{\copyright~CERN on behalf of the \lhcb collaboration,
licence \href{http://creativecommons.org/licenses/by/4.0/}{CC-BY-4.0}.}}
\vspace*{2mm}

\end{titlepage}

\newpage \setcounter{page}{2} \mbox{~}

\cleardoublepage

\renewcommand{\thefootnote}{\arabic{footnote}}
\setcounter{footnote}{0}

\pagestyle{plain} 
\setcounter{page}{1}
\pagenumbering{arabic}

\section{Introduction} \label{sec:Introduction}

\begin{figure}[tb]
\begin{center}
\includegraphics[width=0.48\linewidth]{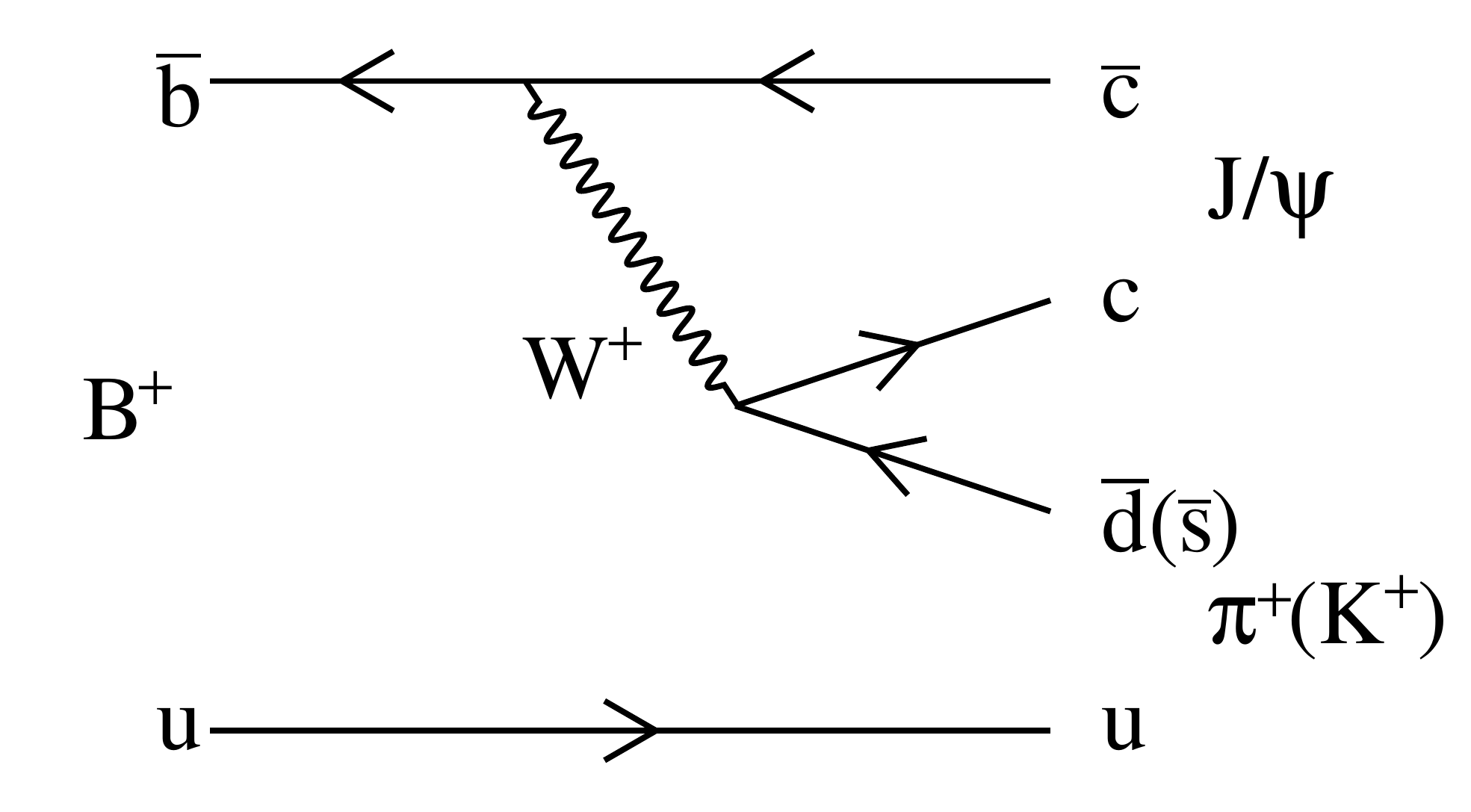}
\includegraphics[width=0.48\linewidth]{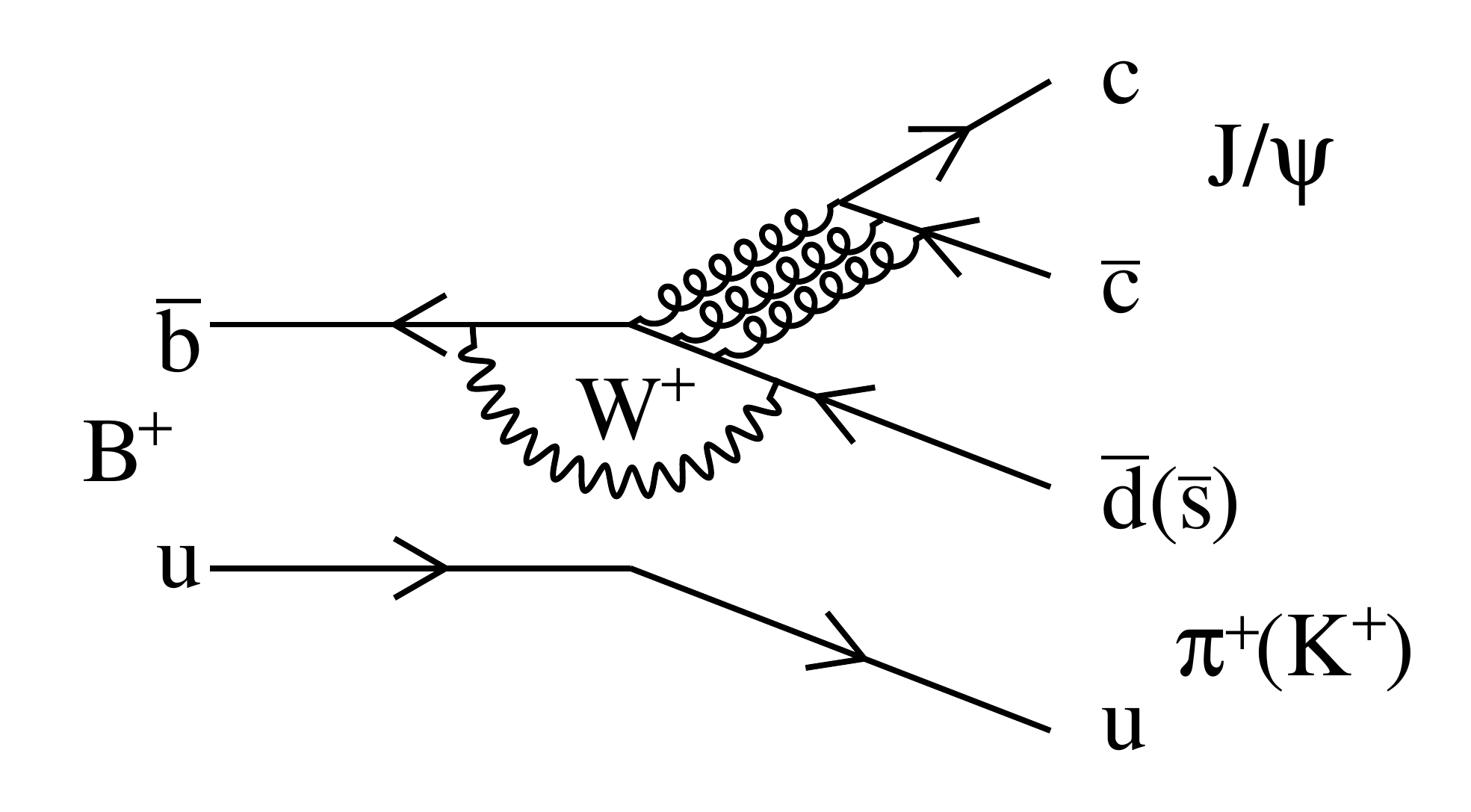}
\vspace*{-0.1cm}
\end{center}
\caption{Feynman diagrams for $\Bpi(\Kp)$ decays at the tree (left) and
one-loop (right) levels.}
\label{fig:Feynman_diagram}
\end{figure}

In the Standard Model, the decay $\Bk$ proceeds via a $b\to c\bar{c}s$ quark
transition\footnote{Unless otherwise specified, the inclusion of
charge-conjugate processes is implied throughout this paper.} and, since this
process is dominated by a Cabibbo-favoured tree diagram, it is expected to
exhibit negligible \CP violation \cite{Dunietz:1993cg}. By contrast, for the
decay $\Bpi$, which proceeds via $b\to c\bar{c}d$, \CP violation up to the
percent level can be generated by interference between the suppressed tree-level
diagram and additional gluonic penguin (loop) diagrams as shown in Fig.\,\ref{fig:Feynman_diagram}.  Measurements of the branching fraction and \CP
asymmetry of the decay $\Bpi$ can provide information about the size of the
penguin-diagram contributions relative to that of the tree diagram. This is
critical for estimating the effects of penguin-diagram contributions in $b\to
c\bar{c}s$ decays on the determination of the \CP violation parameter $\sin 2
\beta$ \cite{Ligeti:2015yma,DeBruyn:2014oga}.

The world average of the branching fraction $\BR(\Bpi)$ is $(4.1\pm 0.4) \times
10^{-4}$~\cite{PDG2016}, with no significant \CP asymmetry observed so far. The
world average value of $\mbox{\ACP(\Bpi)}$, which includes measurements from \belle,
\babar, \dzero and \lhcb
\cite{Abe:2002rc,Aubert:2004pra,Abazov:2013sqa,LHCb-PAPER-2011-024}, is
$(1.0\pm2.8)\times10^{-2}$ \cite{PDG2016}.

In an earlier analysis of a sample of $pp$ collision data corresponding to an
integrated luminosity of $0.37\invfb$ \cite{LHCb-PAPER-2011-024}, \lhcb measured the \CP asymmetry
\mbox{$\ACP(\Bpi) = (0.5\pm2.7\pm1.1)\times10^{-2}$}, as well as the ratio of branching
fractions
\begin{eqnarray}\label{eqn:Rpik}
\rpik\equiv{\BR(\Bpi)\over\BR(\Bk)} =(3.83\,\pm 0.11\pm 0.07) \times 10^{-2}.
\end{eqnarray}
This paper reports an update of the analysis and uses the full $pp$
data sample from the \lhc Run 1, corresponding to $1\invfb$ collected at a
centre-of-mass energy of $7\tev$ and $2\invfb$ at 8\tev, and measures $\rpik$
and $\apik\equiv \ACP(\Bpi)-\ACP(\Bk)$, where these two decays are reconstructed
using the dimuon decay mode of the $\jpsi$ meson. The result for $\apik$ is combined
with the $\ACP(\Bk)$ measurement from another \lhcb analysis
\cite{LHCb-PAPER-2016-054} to obtain $\ACP(\Bpi)$.

\section{Detector and simulation}
\label{sec:Detector}

The \lhcb detector~\cite{Alves:2008zz,LHCb-DP-2014-002} is a single-arm forward
spectrometer covering the \mbox{pseudorapidity} range $2<\eta <5$, designed for
the study of particles containing \bquark or \cquark quarks. The detector
includes a high-precision tracking system consisting of a silicon-strip vertex
detector surrounding the $pp$ interaction region, a
large-area silicon-strip detector located upstream of a dipole magnet with a
bending power of about $4{\mathrm{\,Tm}}$, and three stations of silicon-strip
detectors and straw drift tubes placed downstream of the
magnet.  The tracking system provides a measurement of momentum, \ptot, of
charged particles with a relative uncertainty that varies from 0.5\% at low
momentum to 1.0\% at 200\gevc.  The minimum distance of a track to a primary
vertex (PV), the impact parameter, is measured with a resolution of
$(15+29/\pt)\mum$, where \pt is the component of the momentum transverse to the
beam, in\,\gevc.  Different types of charged hadrons are distinguished using
information from two ring-imaging Cherenkov detectors.
Photons, electrons and hadrons are identified by a calorimeter system consisting
of scintillating-pad and preshower detectors, an electromagnetic calorimeter and
a hadronic calorimeter. Muons are identified by a system composed of alternating
layers of iron and multiwire proportional chambers.  The
online event selection is performed by a trigger~\cite{LHCb-DP-2012-004}, which
consists of a hardware stage, based on information from the calorimeter and muon
systems, followed by a software stage, which applies a full event
reconstruction.  

In this analysis, the hardware trigger decision is required to be caused by at
least one high-$\pt$ track that is consistent with being a muon.  In the
software trigger, two well-reconstructed muons with opposite charge are required
to form a good-quality vertex and to have an invariant mass consistent with that
of the $\jpsi$ meson \cite{PDG2016}.  The trigger also requires a significant
displacement between the $\jpsi$ vertex and the associated PV of the $pp$
collision.

In the simulation, $pp$ collisions are generated using
\pythia~\cite{Sjostrand:2006za,Sjostrand:2007gs} with a specific \lhcb
configuration~\cite{LHCb-PROC-2010-056}.  Decays of hadronic particles are
described by \evtgen~\cite{Lange:2001uf}, in which final-state radiation is
generated using \photos~\cite{Golonka:2005pn}. The interaction of the generated
particles with the detector, and its response, are implemented using the \geant
toolkit~\cite{Allison:2006ve, *Agostinelli:2002hh} as described in
Ref.~\cite{LHCb-PROC-2011-006}.

\section{Event selection}
\label{sec:eventselection}

The same criteria are used to select $\Bpi$ and $\Bk$ decays, except for those
related to the identification of the final-state hadrons, and consist of a loose
preselection followed by a multivariate selection.  In the preselection, all
three final-state tracks are required to be of good quality and within a
fiducial region of the detector acceptance that excludes areas with large
asymmetries in the detection efficiencies.

The $\jpsi$ candidates are formed from two oppositely charged particles with
$\pt$ greater than $550\mevc$, identified as muons and consistent with
originating from a common vertex  but inconsistent  with originating from any
PV.  The invariant mass of the $\mup \mun$ pair is required to be  within
$^{+43}_{-48}\mevcc$ of the known $\jpsi$ mass \cite{PDG2016}, then constrained
to that value in subsequent stages of the reconstruction. The $\Bp$ candidates
are formed by combining each $\jpsi$ candidate with a hadron candidate that has
$\pt$ greater than $1\gevc$ and $p$ greater than $5\gevc$ and forms a common
vertex with the $\jpsi$. Both the kaon and pion mass hypotheses of the hadron
candidates are kept. Each reconstructed $\Bp$ candidate is required to be
consistent with  originating from a PV. The vector from the corresponding PV to
the decay vertex of the $\Bp$ is required to be closely aligned with the
momentum vector of the $\Bp$ candidate: the opening angle $\phi$ between them
must satisfy $\cos\phi>0.999$.  To ensure a clean separation between the $\Bpi$
and $\Bk$ mass peaks in the $\jpsi\pip$  mass spectrum, the decay angle
$\theta_h$, defined as the angle between the momentum of the kaon or pion in the
$\Bp$ rest frame and the $\Bp$ momentum in the laboratory frame, is required to
satisfy  $\cos\theta_h<0$ \cite{LHCb-PAPER-2011-024}. 

The $\Bpi$ and $\Bk$ candidates passing the preselection are filtered using the
output of a boosted decision tree (BDT) \cite{Breiman,AdaBoost} to further suppress
combinatorial background. The BDT uses kinematic and topological
variables to discriminate between signal and background.  These include the
impact parameters of the final-state tracks with respect to the PV, as well as
those of the $\jpsi$ and the $\Bp$ candidates, the $\pt$ of the final-state
hadron and the $\jpsi$ and $\Bp$ candidates, and the decay-length and
vertex-fit $\chi^2$ of the $\Bp$ candidate. Given the similarity of their
kinematic distributions, the same BDT classifier is used to
select both decays. The
BDT is trained using a simulated sample of $\Bpi$ decays and a background sample
consisting of candidates from the data sample passing the $\Bpi$ preselection
with invariant mass in the range 5500--5700\mevcc. 

Particle identification (PID) criteria are applied to select pion and kaon
candidates, with the two hypotheses being mutually exclusive. The requirements
on the BDT response and PID are chosen to maximise the figure of merit for the
decay $\Bpi$, defined as $N_{\pi}/\sqrt{N_{\rm tot}}$, where $N_{\rm tot}$ is
the total number of $\Bpi$ candidates within $\pm 3$ times the mass resolution
around the known $\Bp$ mass.  Here $N_{\pi}$ refers to the $\Bpi$ signal yield
and is estimated to be $(N_{\rm tot}-N_{\rm comb})/(1+1/(r_{\rm eff}\rpik))$,
where the value of $\rpik$  is given in Eq.  \ref{eqn:Rpik}, $N_{\rm comb}$ is
the number of combinatorial background events in the $\Bpi$ signal region
extrapolated from the region 5340--5580\mevcc passing the PID selection, and
$r_{\rm eff}$ is the ratio of the efficiencies for $\Bpi$ and $\Bk$
events to pass the $\Bpi$  selection and fall in the signal window, estimated
from simulation. After this optimisation, the BDT rejects more than $85\%$ of
the combinatorial background and retains around $92\%$ of $\Bu\to\jpsi h^+$
events, where $h=\pi$, $K$.  The particle identification requirement has an
efficiency of about $97\%$ for $\Bpi$ and $69\%$ for $\Bk$. The fraction of
events in which more than one candidate passes the selection is negligible.

\section{Signal yield determination}%
\label{sec:signaldetermination}

\begin{figure}[tb]
\begin{center}
\includegraphics[width=0.48\linewidth]{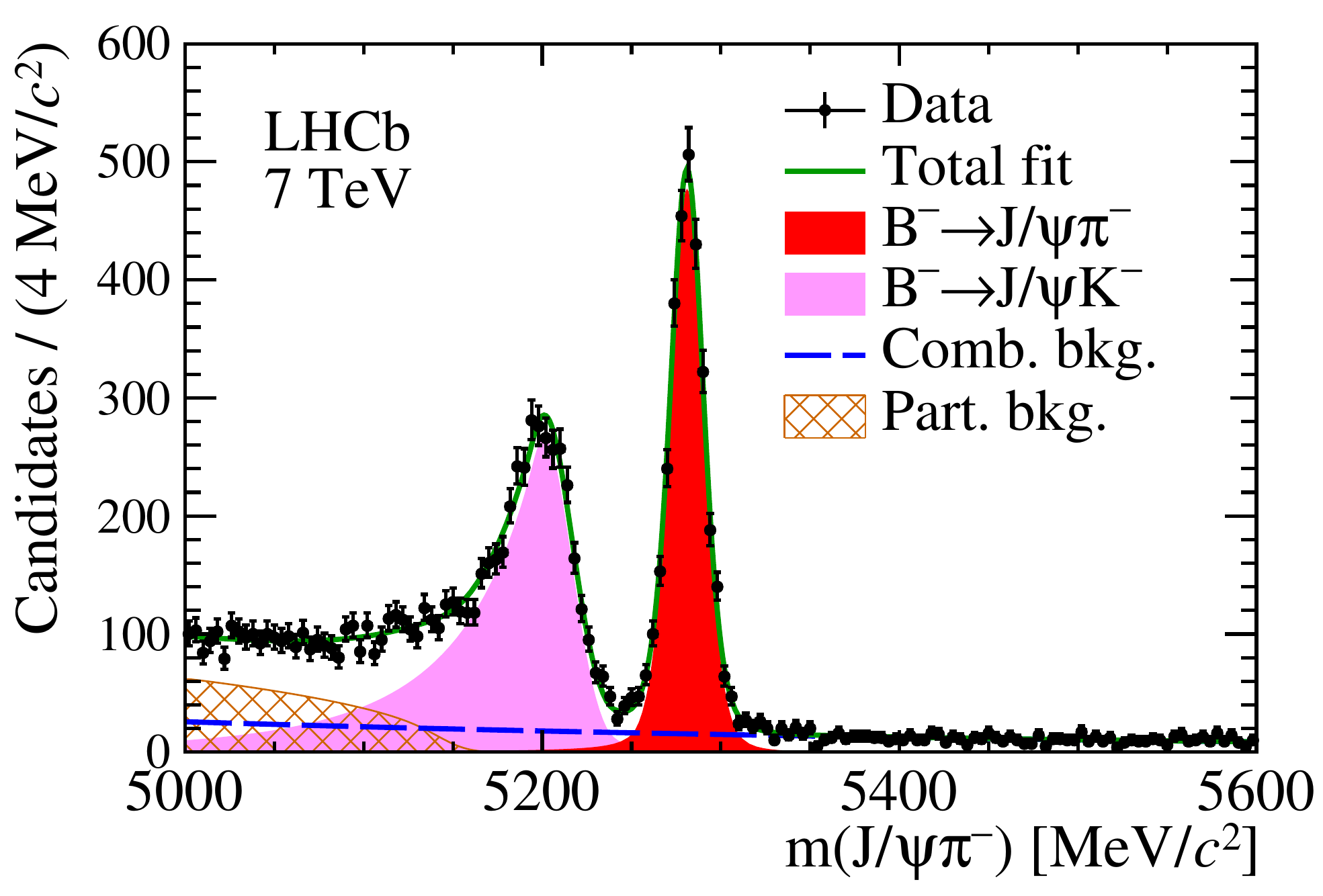}
\includegraphics[width=0.48\linewidth]{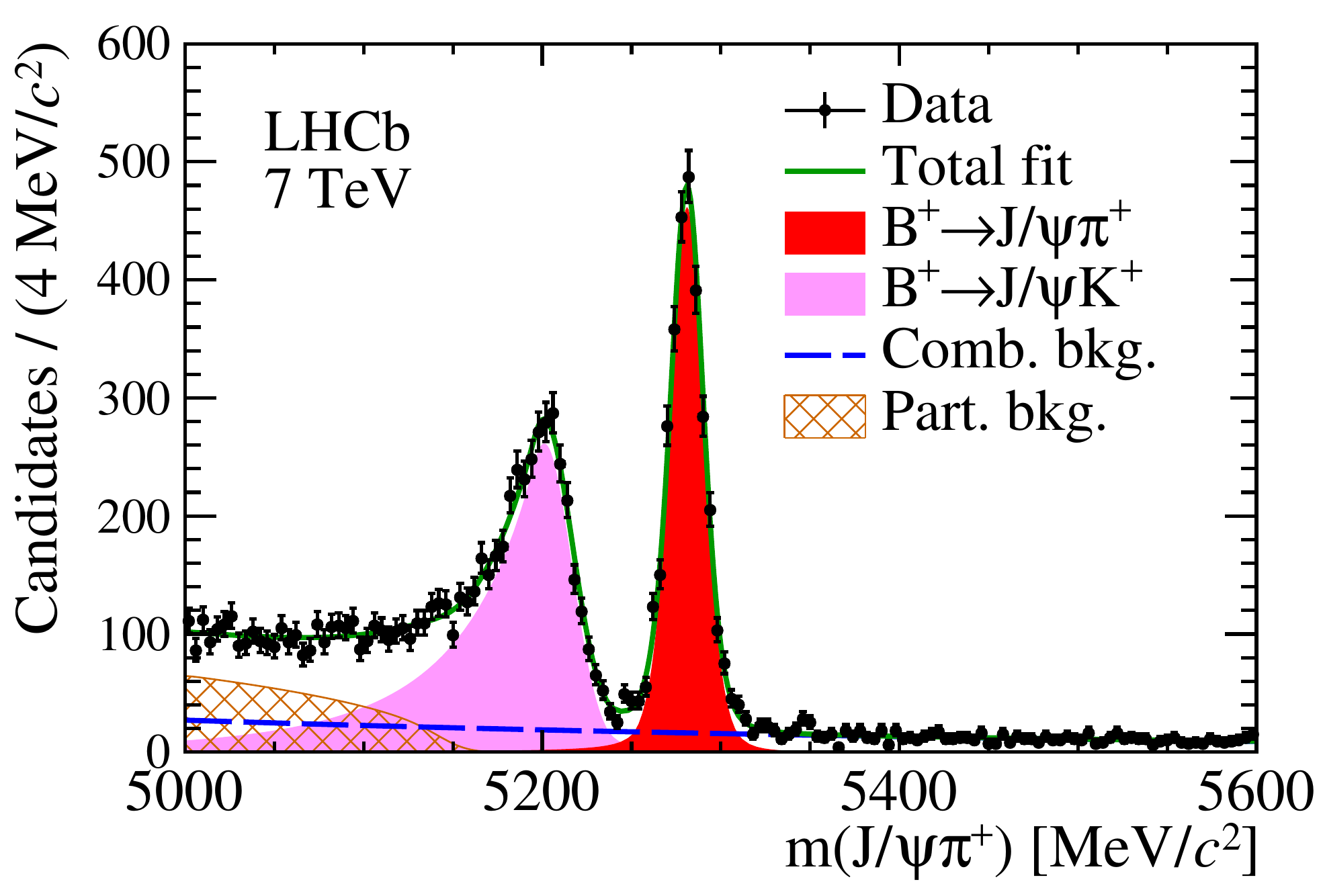}
\includegraphics[width=0.48\linewidth]{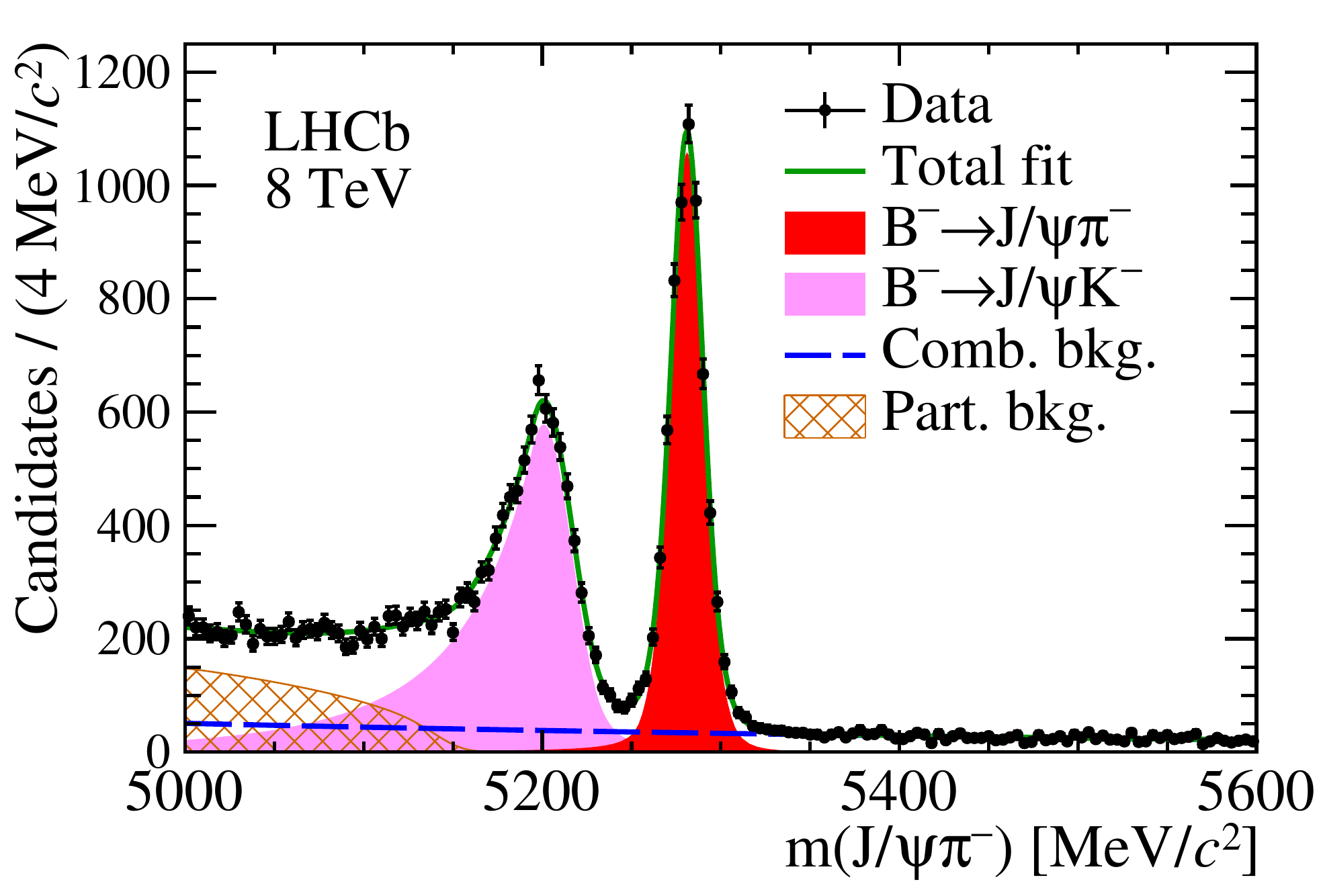}
\includegraphics[width=0.48\linewidth]{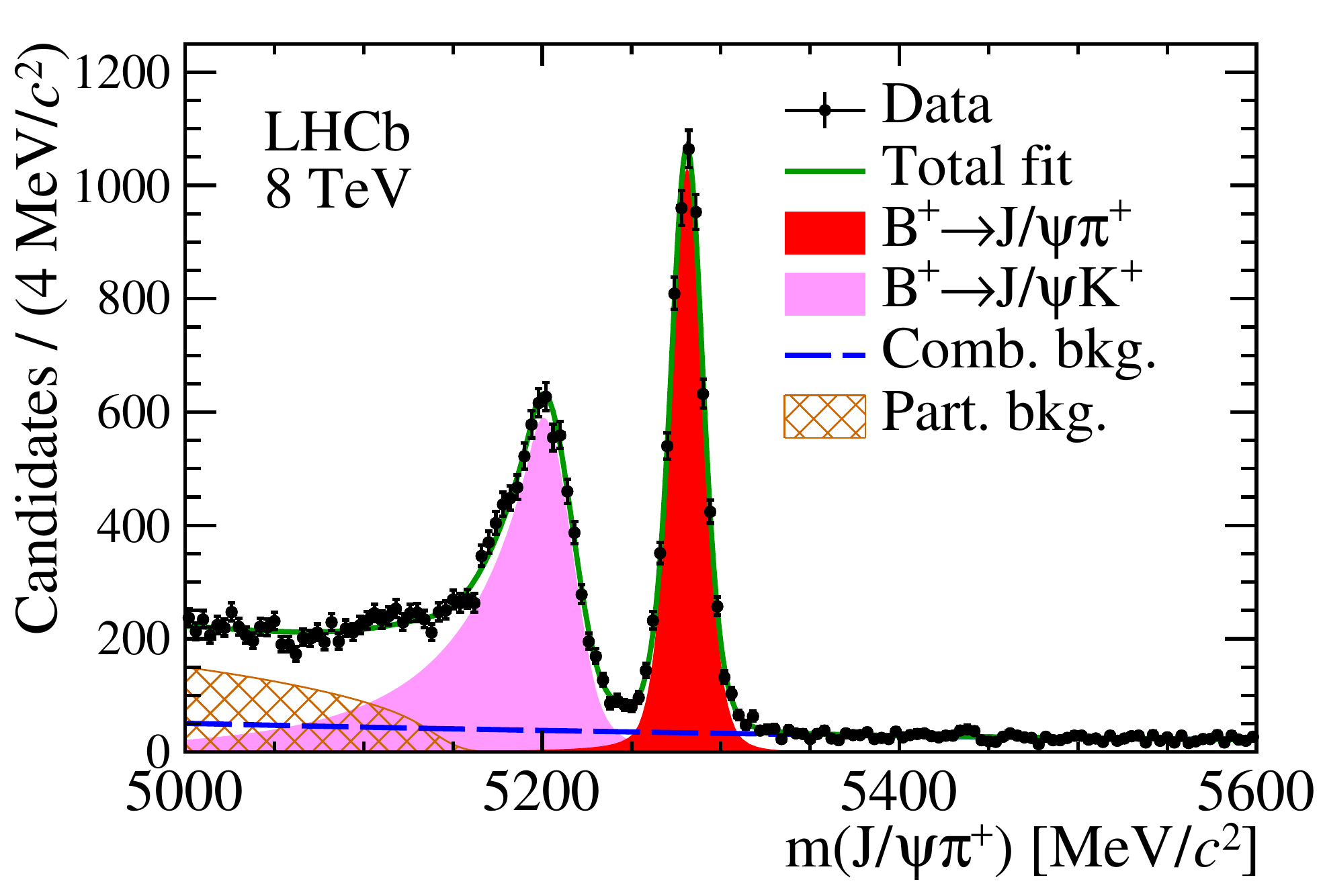}
\vspace*{-0.1cm}
\end{center}
\caption{Invariant mass distributions of (left)
$\Bub\to\jpsi\pim$ and (right) $\Bu\to\jpsi\pip$ candidates with the result of
the fit superimposed, for data collected at (top) 7 TeV and (bottom) 8 TeV.
}\label{fig:fit_B2JpsiPi}
\end{figure}

\begin{figure}[tb] \begin{center}
\includegraphics[width=0.48\linewidth]{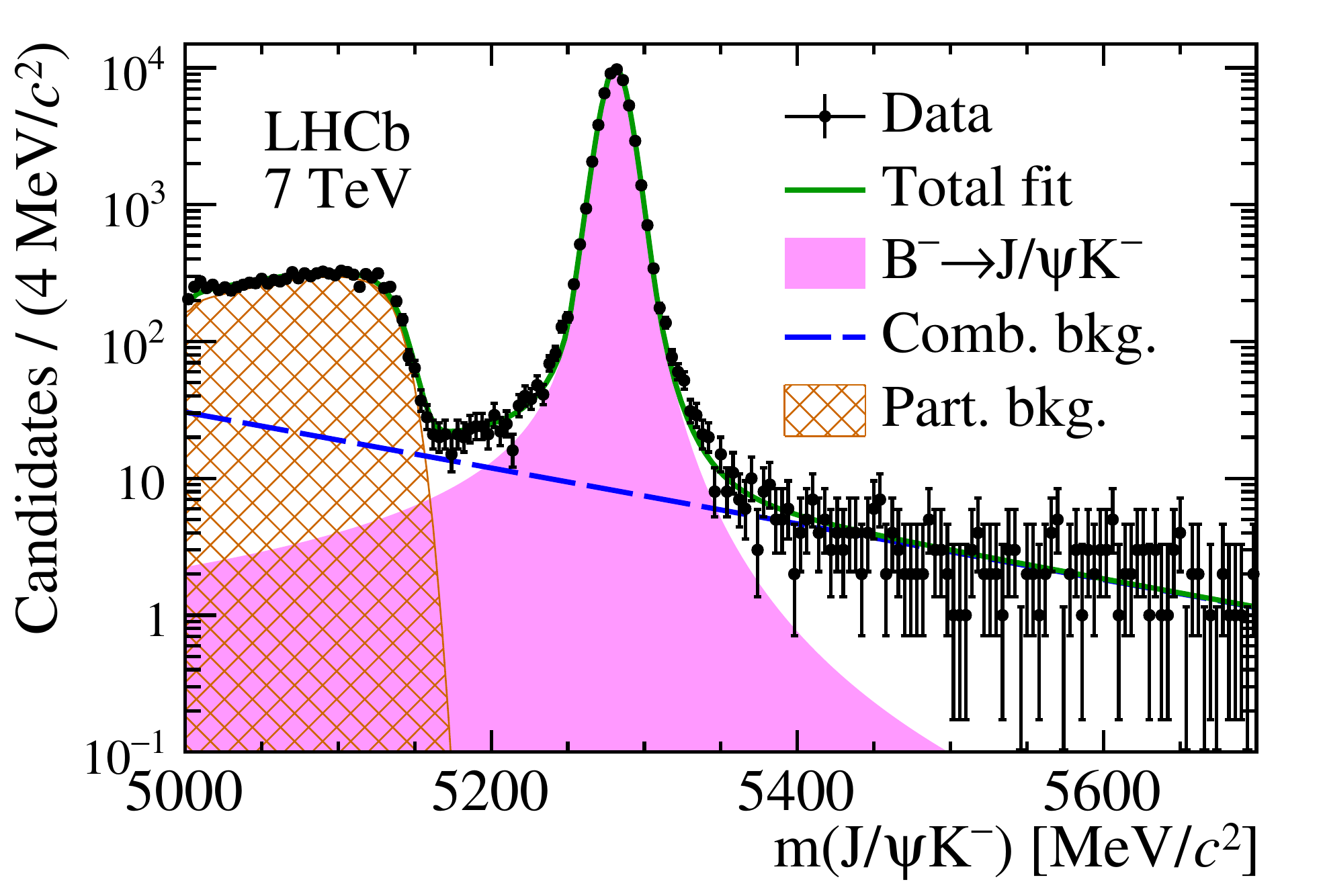}
\includegraphics[width=0.48\linewidth]{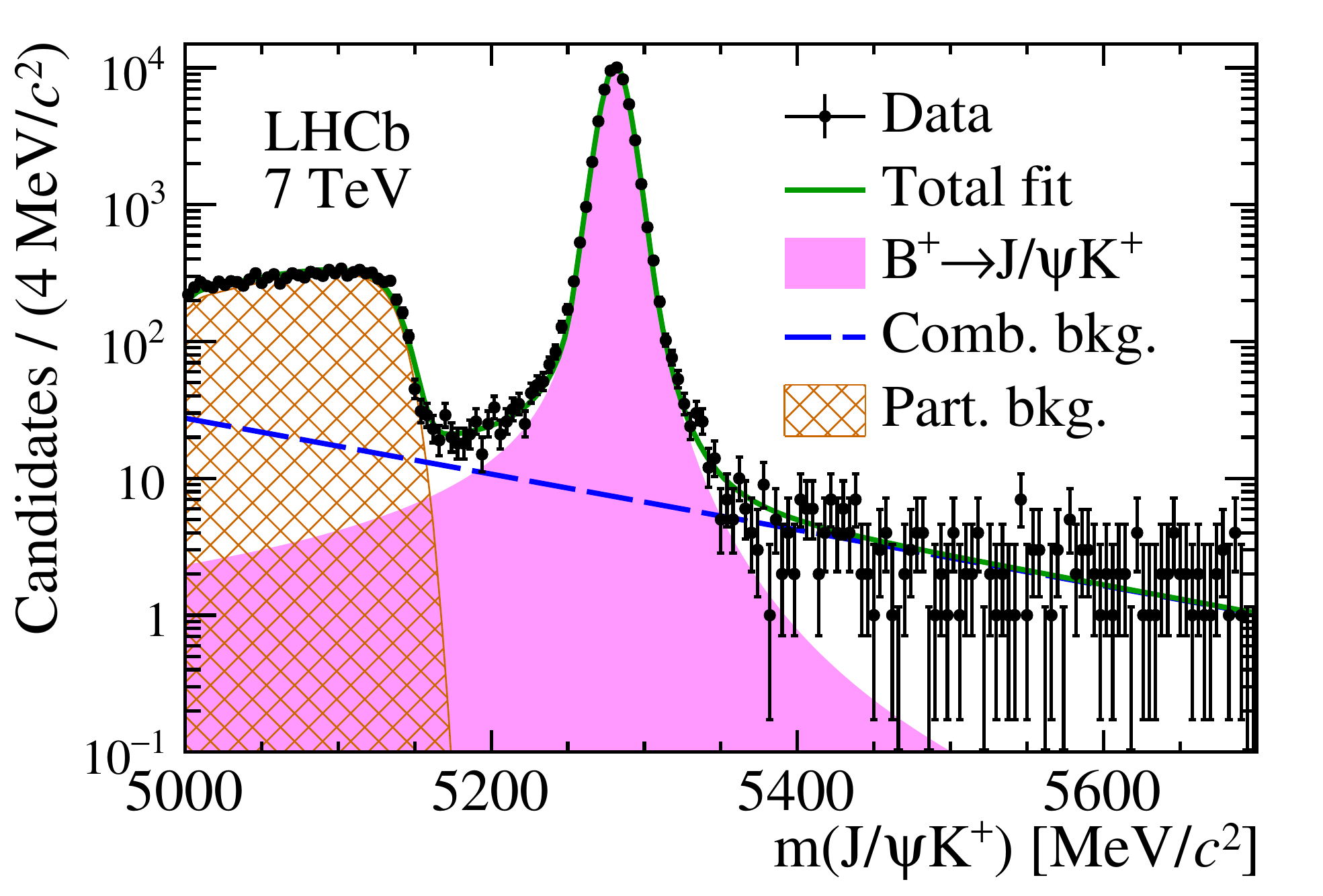}
\includegraphics[width=0.48\linewidth]{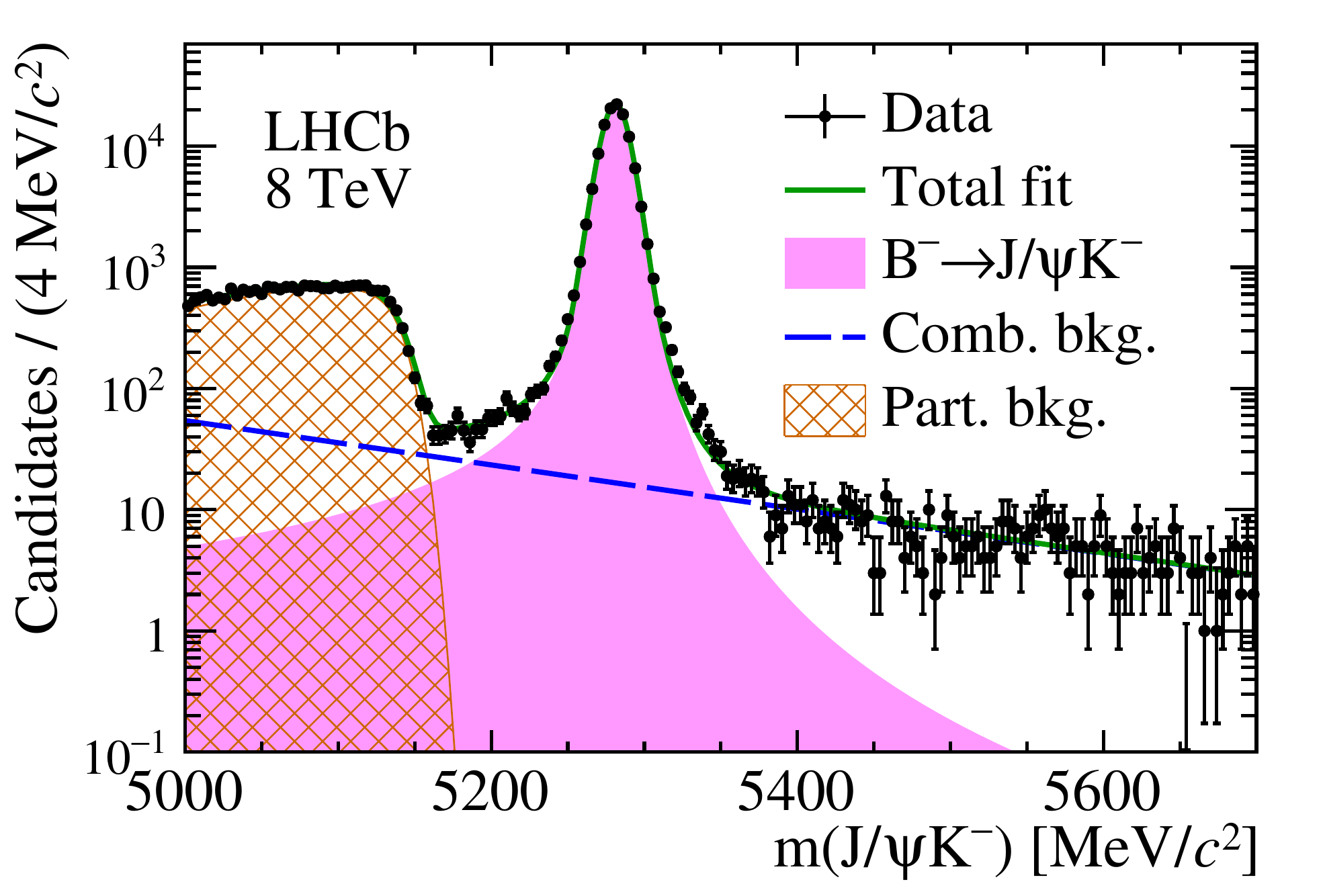}
\includegraphics[width=0.48\linewidth]{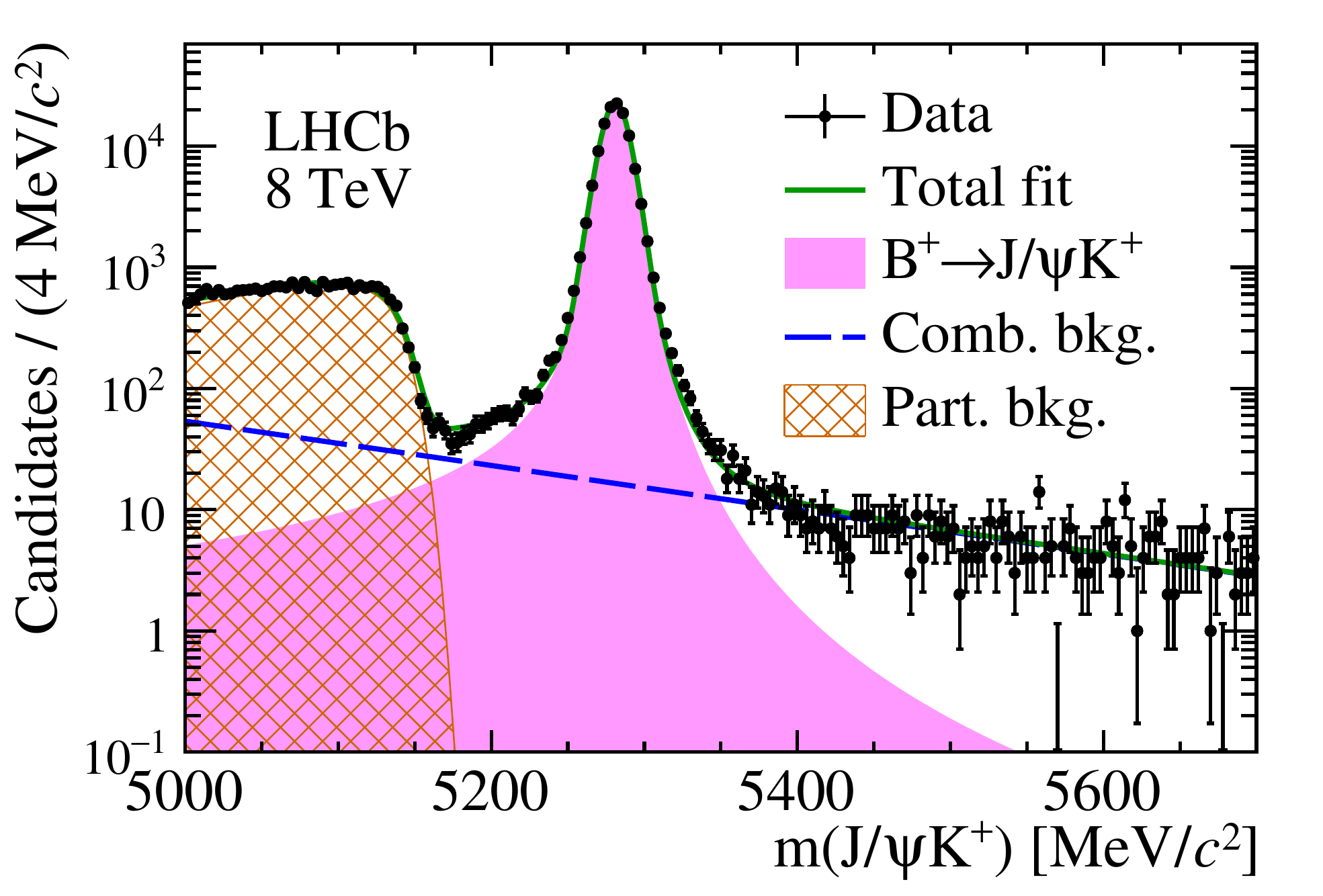}
\vspace*{-0.1cm} \end{center} \caption{Invariant mass distributions of (left)
$\Bub\to\jpsi\Km$ and (right) $\Bu\to\jpsi\Kp$ candidates with the result of the
fit superimposed, for data collected at (top) 7 TeV and (bottom) 8 TeV, where
the $\Bpm\to\jpsi\pipm$ contributions are neglected.
}\label{fig:fit_B2JpsiK} \end{figure}

The signal yields $N_{\jpsi h}$ and raw charge asymmetries $A^{\rm raw}_{\jpsi
h}$ of the two decay modes are determined from independent unbinned extended
maximum likelihood fits to the invariant mass distributions of $\Bu\to\jpsi
h^+$ and $\Bm\to\jpsi h^-$. Denoting the signal yield for $B^\pm\to\jpsi
h^\pm$ by $N_{\jpsi h^\pm}$, $N_{\jpsi h}$ is the sum of $\Bm\to\jpsi\pim$ and $\Bpi$, and $A^{\rm raw}_{\jpsi h}$ is defined as 
\begin{eqnarray} 
A^{\rm raw}_{\jpsi h}={N_{\jpsi h^-}-N_{\jpsi h^+}\over N_{\jpsi h^-}+N_{\jpsi h^+}}.
\end{eqnarray}
The fits use $\Bpi$ candidates in the range 5000--5600\mevcc and $\Bk$
candidates in the range 5000--5700\mevcc.  The $\Bp$ and $\Bm$ samples are
fitted simultaneously, as shown in Figs. \ref{fig:fit_B2JpsiPi} and
\ref{fig:fit_B2JpsiK}.  Table \ref{tab:signal_yields} summarizes the fit results
for the parameters of interest. In each
fit, the signal shape is modelled by a Hypatia function \cite{Santos:2013gra}.
The most probable value and the resolution of the Hypatia function are allowed to vary in the
fit, while the tail parameters are fixed to values determined from fits to
simulated events. The hadron misidentification background in the $\Bpi$ sample,
arising from $\Bk$ decays in which the kaon is misidentified as a pion, is
described by a double-sided Crystal Ball (DSCB) function whose parameters,
except for the most probable value and the core width, are
fixed to values determined from fits to simulated events. The misidentification background
due to $\Bpi$ decays in which the pion is misidentified as a kaon is neglected
in the baseline fit; a systematic uncertainty due to this assumption is
assigned, as discussed in Sec. \ref{sec:systematic}. The combinatorial
background is modelled by an exponential function whose shape parameter is left free
in the fit. The background due to partially reconstructed $B$-meson decays such
as $B\to\jpsi h\pi$ is described by an ARGUS function \cite{Albrecht:1990am}
convolved with a Gaussian function, with all parameters allowed to vary in the fit. Contributions from the highly suppressed
$\Bu\to\Kp\mumu$ \cite{PDG2016} and $\Bu\to\pip\mumu$ \cite{LHCb-PAPER-2012-020}
decays are negligible. 

\begin{table}[tb]
\caption{Signal yields and raw charge asymmetries determined
   from the fits, which are described in the text. The uncertainties are
statistical.}\label{tab:signal_yields}
\begin{center}
\begin{tabular}{lr@{ $\pm$ }rr@{ $\pm$ }r}
   & \multicolumn{2}{c}{7\tev}\phantom{000} &
   \multicolumn{2}{c}{8\tev}\phantom{000} \\ \hline
   $N_{\jpsi\pi}$ & 6011 & 89\phantom{$0.0\times10^{-2}$} & 13\,103 &
   130\phantom{$.0\times10^{-2}$} \\
   $N_{\jpsi\kaon}$ & 107\,783 & 332\phantom{$.0\times10^{-2}$} & 243\,119 & 499\phantom{$.0\times10^{-2}$} \\
   $A^{\rm raw}_{\jpsi\pion}$ & $(1.64$ & $1.39)\times10^{-2}$ & $(1.35$ &
   $0.94)\times10^{-2}$ \\
   $A^{\rm raw}_{\jpsi\kaon}$ & $(-1.65$ & $0.31)\times10^{-2}$ & $(-1.27$ &
   $0.20)\times10^{-2}$ \\
\end{tabular}
\end{center}
\end{table}

\section{Efficiency corrections}
\label{sec:efficiency}

The ratio of the $\Bpi$ and $\Bk$ branching fractions is measured separately for
the 7 and 8\tev samples, and is calculated as
\begin{eqnarray}\label{eqn:branching_fraction}
\rpik
   ={{N_{\jpsi\pi}}\over{N_{\jpsi\kaon}}}
   \times{\eps_{\jpsi\kaon}\over\eps_{\jpsi\pion}},
\end{eqnarray}
where $\eps_{\jpsi\pion}$ and $\eps_{\jpsi\kaon}$ denote the total efficiencies
of selecting the two modes, each taking into account the geometrical acceptance
of the detector, the trigger, the reconstruction and preselection, the hadron
PID, the BDT selection and the fiducial selection.  The hadron PID efficiencies
are determined using  $\Dstarp\to\Dz(\to\Km\pip)\pi^+$ calibration data
\cite{Anderlini:2202412}.  Kaons and pions in the calibration samples are
weighted to reproduce  the momentum and pseudorapidity distributions of those
from $\Bk$ and $\Bpi$  decays.  All other  efficiencies are estimated using
simulated signal events.  The simulated events are weighted such that their
kinematic distributions match those of the background-subtracted data, which is
obtained using the \sPlot technique \cite{Pivk:2004ty}. The efficiency ratio,
$\eps_{\jpsi\pi}/\eps_{\jpsi\kaon}$, is estimated to be $1.43\pm0.01$ for the
$7\tev$ data and $1.42\pm0.01$ for $8\tev$, with the difference from unity being
mainly due to the PID selections for the two decays.

The difference in \CP asymmetries of $\Bpi$ and $\Bk$ is calculated as
\begin{eqnarray}\label{eqn:asymmetry}
\Delta \ACP & = & \Delta A^{\rm raw} -
\Delta A^{\rm eff}\;, \nonumber\\
\Delta A^{\rm raw} & \equiv& A^{\rm
raw}_{\jpsi\pion}-A^{\rm raw}_{\jpsi\kaon}\;, \nonumber\\
\Delta A^{\rm eff} &
\equiv& A^{\rm eff}_{\jpsi\pion}-A^{\rm eff}_{\jpsi\kaon}\;,
\end{eqnarray}
where $A^{\rm eff}_{\jpsi\pion}$ and $A^{\rm eff}_{\jpsi\kaon}$ are the
efficiency asymmetries between $B^-$ and $B^+$ decays. The asymmetry difference
$\Delta A^{\rm eff}$ arises from the particle detection
efficiency, hadron PID, BDT selection and fiducial selection. The main sources
of asymmetry are the detection efficiency and hadron PID, as described below.

The PID efficiency asymmetries of $\Bpi$ and $\Bk$ are estimated separately
using the $\Dstarp\to\Dz(\to\Km\pip)\pi^+$ calibration sample mentioned above, and
their difference is taken as a contribution to $\Delta A^{\rm eff}$.  The
average detection asymmetry between $\pim$ and $\pip$ in $\Bpi$ is denoted
$A^{\rm det}_{\pi}$, and that between $\Km$ and $\Kp$ in $\Bk$ is likewise
denoted $A^{\rm det}_{K}$. Following the method in
Ref.\cite{LHCb-PAPER-2014-069}, the difference $A^{\rm det}_{\pi} - A^{\rm
det}_{K}$ can be approximated by the combined  detection asymmetry between
$\pim\Kp$ and $\pip\Km$, denoted $A^{\rm det}_{\pi \Kbar}$, which is
calculated as
\begin{eqnarray}
A^{\rm det}_{\pi} - A^{\rm det}_{K} \approx
A^{\rm det}_{\pi \Kbar} =A^{\rm raw}_{\Dm\to\Kp\pim\pim}-A^{\rm
raw}_{\Dm\to\KS\pim}+A^{\rm det}_{\KS}\;. 
\end{eqnarray}
Here $A^{\rm
raw}_{\Dm\to\Kp\pim\pim}$ and $A^{\rm raw}_{\Dm\to\KS\pim}$ are the raw charge
asymmetries measured in the decays $\Dm\to\Kp\pim\pim$ and $\Dm\to\KS\pim$. The
$\Dmp$ production asymmetry cancels in the difference between the two raw
asymmetries, and the \CP asymmetries in Cabibbo-favoured charm decays are
assumed to be negligible.  The $\Dm\to\Kp\pim\pim$ decays are weighted to match
the distributions of $\pt$ and rapidity ($y$) of kaons in the $\Bk$ decays. The
$\Dm\to\KS\pim$ decays are then weighted to match the kinematic distributions of
the $\Dm\to\Kp\pim\pim$ sample such that the $\pt$ and $y$ distributions of the
$\Dm$ agree between the two channels, as do the $\pt$ distributions of the
$\pim$ (with one pion chosen at random in the case of $\Dm\to\Kp\pim\pim$). The
term $A^{\rm det}_{\KS}$ is a small correction for the effects of \CP violation
in $\Kz$--$\Kzb$ mixing and the different interaction cross-sections of $\Kz$
and $\Kzb$ with the detector material \cite{LHCb-PAPER-2014-013}.  The asymmetry
$A^{\rm det}_{\pi\Kbar}$ is evaluated to be $(1.10\pm0.22) \times 10^{-2}$ and
$(0.77\pm0.10)\times 10^{-2}$ for the 7 and 8 $\tev$ data, respectively.  The
overall difference in efficiency asymmetry, $\Delta A^{\rm eff}$, is estimated
to be $(1.37\pm0.56)\times10^{-2}$ for the $7\tev$ data, and
$(0.84\pm0.43)\times 10 ^{-2}$ for $8\tev$.

\section{Systematic uncertainties}
\label{sec:systematic}

The data-taking conditions were different for the $7$ and $8\tev$ data,
and therefore the systematic uncertainties, summarised in Table
\ref{tab:sys_summary}, are computed separately for the two samples. The relative
uncertainties are 
quoted for the $\rpik$ measurement and absolute uncertainties are quoted for the
$\Delta\ACP$ measurement.  The systematic uncertainties can be divided into two
groups, either associated with the mass fit or with the efficiency. For each
systematic uncertainty associated with the mass fit, a fit with an alternative model is performed and the
differences in the mean values of $\rpik$ and $\Delta\ACP$ are taken as the
corresponding systematic uncertainties. The alternative fits are performed
with the same sets of parameters floating or fixed as in nominal fit. In each case, the uncertainties are quoted
separately for the $7$ and $8\tev$ data.

The baseline signal model is a Hypatia function. Changing this to a histogram
representing the simulated signal mass distribution convolved with a Gaussian
function, to correct for mismatch in resolution between data and simulation,
leads to relative uncertainties of $0.39\%$ and $0.25\%$ for $\rpik$ for the 7
and 8\tev data
and absolute uncertainties of $0.03 \times 10 ^{-2}$ and less than
$0.01\times10^{-2}$
for $\Delta\ACP$.  
 
The baseline model for the misidentification background in the $\Bpi$ sample  is
a DSCB function with tail parameters  obtained from the simulation.  Alternative
models are constructed by  varying the tail parameter values to match those
expected for
different  pion selection requirements, or by using a histogram
convolved with a Gaussian function as was done for the signal model. The results
from different alternative models are summed in
quadrature. The resulting relative systematic
uncertainties on $\rpik$ are  $0.44\%$ and $0.38\%$, and the estimated
systematic uncertainties on $\Delta\ACP$ are $0.01 \times 10 ^{-2}$ and
$0.02 \times 10 ^{-2}$.

The most probable values and the resolution parameters of the signal and misidentification
background models are assumed to be the same for $\Bp$ and $\Bm$ decays in the
baseline fits. Treating the parameters separately for $\Bp$ and
$\Bm$  decays leads to differences (taken as estimates of the associated
uncertainties) of $0.04 \times 10 ^{-2}$ and $0.05 \times
10 ^{-2}$ for $\Delta\ACP$ and $0.04\%$ and $0.02\%$ for $\rpik$.

The baseline model for the combinatorial background is an exponential function.
Adding a linear component to this model shifts $\rpik$ by $0.52\%$ and $0.20\%$,
and changes $\Delta\ACP$ by $0.04 \times 10 ^{-2}$ and $0.01 \times 10
^{-2}$.

The baseline fits are performed in  mass ranges above $5000\mevcc$, where
contamination from the partially  reconstructed background is expected up to
$5150\mevcc$.  The alternative fits are performed in narrower ranges starting
from $5150\mevcc$, where  partially  reconstructed background can be neglected.
The  value of $\rpik$ is found to change by $0.20\%$ and $0.33\%$, and that
of $\Delta\ACP$  by $0.04 \times 10 ^{-2}$ and $0.01 \times 10
^{-2}$. Systematic uncertainties equal to these shifts are assigned.

The PID efficiencies are calibrated using $\Dstarp\to\Dz(\to\Km\pip)\pip$
decays selected without applying hadron PID requirements. The efficiency depends
on the momentum and pseudorapidity of the track and the track multiplicity in the event, and the
calibration is therefore done in bins of those variables. The choice of binning
necessarily involves a compromise between the granularity and statistical
uncertainty of individual bins. Systematic uncertainties due to the limited number
of kinematic bins are evaluated by doubling or halving the number of bins and
recalculating the average efficiencies. The resulting deviations from the
baseline results are taken as the systematic uncertainties: $0.39\%$ and
$0.46\%$ for $\rpik$, and $0.06 \times 10^{-2}$ and $0.01 \times 10^{-2}$ for
$\Delta\ACP$.

The ratio of BDT efficiencies of the decays $\Bpi$ and $\Bk$ is estimated with simulated
samples of signal events, which are weighted to remove differences in the
distributions of the BDT input variables between the simulation and data. Relative
systematic uncertainties of $0.01\%$ and $0.02\%$ are assigned to $\rpik$, to
account for statistical uncertainties on the weights used in the efficiency
calculation.

The ratio of trigger efficiencies of the decays $\Bpi$ and $\Bk$ is determined
from
simulation and validated with a control sample of $\jpsi\to\mumu$ decays
\cite{LHCb-DP-2012-004}.  Relative differences of $0.33\%$ and $0.38\%$ are
found between the values of this ratio estimated with data and with simulation, which
are taken as the corresponding systematic uncertainties on $\rpik$.

Samples of $\Dp$ decays are used to determine the difference between the kaon and
pion detection efficiency asymmetries. However, the kinematic distributions of
the pions and kaons in the $\Dp$ samples may differ from those of the signal
$\Bu\to\jpsi h^+$ samples, and the efficiency asymmetries may vary with the particle
kinematics. To assess the scale of this effect, samples of
$\Dp\to\Km\pip\pip$ events are weighted such that the distribution of the
momentum of the kaon
matches that of $\Bk$, leading to a pion detection asymmetry of
$0.12\times10^{-2}$ for both 7 and $8\tev$ data. This is taken as a systematic uncertainty.

The production asymmetry of $\Bu$ mesons is a function of the $\Bu$
kinematics. This dependence cancels in the observables considered, provided that
$\Bpi$ and $\Bk$ decays have the same kinematic distributions. Good agreement is
found between the $\pt$ distributions of the decays $\Bpi$ and $\Bk$, but not for the
rapidity distributions. The deviations of the $\Bp$ production asymmetry with
and without the weights that match the rapidity distribution in the $\Bpi$
sample to that of the $\Bk$ decay, are $0.02 \times 10^{-2}$ and $0.04 \times
10^{-2}$, which are taken as the systematic uncertainties on $\Delta\ACP$.

A systematic uncertainty of $0.03\%$ on $\rpik$ is assigned to account for
imperfect simulation of hadron interactions in the detector, determined
from the known interaction cross-sections for pions and kaons and 
assuming an uncertainty of $10\%$ in the material budget of the detector.
Summing all of the above contributions in quadrature, the relative systematic
uncertainty on $\rpik$ is $1.01\%$ for the $7\tev$ sample and $0.83\%$ for
$8\tev$ and the absolute uncertainty on $\Delta\ACP$ is $0.15 \times
10^{-2}$ for $7\tev$ and $0.14\times10^{-2}$ for $8\tev$.

\begin{table}[tb]
 \caption{Relative systematic uncertainties ($\%$) for $\rpik$ and 
absolute systematic uncertainties ($\times10^{-2}$) for $\Delta\ACP$. The
uncertainties are quoted separately for the 7 and 8 TeV data. The
dashes indicate negligible uncertainties (zero after rounding to two decimal
places).}
 \label{tab:sys_summary}
 \begin{center}
 \begin{tabular}{lcccc}
 \hline
 Sources & $\rpik$ (7\tev) & $\rpik$ (8\tev) & $\Delta\ACP$ (7\tev) &
 $\Delta\ACP$ (8\tev) \\
 & [\%] & [\%] & $[\times10^{-2}]$ & $[\times10^{-2}]$ \\
 \hline
 Signal model
 & 0.39 & 0.25 & 0.03 & --  \\
 Mis-ID background
 & 0.44 & 0.38 & 0.01 & 0.02 \\
 $\Bpm$ parameters
 & 0.04 & 0.02 & 0.04 & 0.05 \\
 Comb. background
 & 0.52 & 0.20 & 0.04 & 0.01 \\
 Part. reco.  background
 & 0.20 & 0.33 & 0.04 & 0.01 \\
 PID efficiency 
 & 0.39 & 0.46 & 0.06 & 0.01 \\
 BDT efficiency
 & 0.01 & 0.02 & -- & -- \\
 Trigger efficiency
 & 0.33 & 0.38 & -- & -- \\
 Detection asymmetry
 & -- & -- & 0.12 & 0.12 \\
 $\Bpm$ prod. asymmetry 
 & -- & -- & 0.02 & 0.04 \\
 $K/\pi$ interaction 
 & $0.03$  & $0.03$ & --  & -- \\
 \hline
 Total 
 & 1.01 & 0.83 & 0.15 & 0.14 \\
 \hline
 \end{tabular}
\end{center}
\end{table}

\section{Results and conclusion}
\label{sec:results}

Using the estimated signal yields, efficiency ratios, raw charge asymmetries and
efficiency asymmetries, the ratio of branching fractions and difference in \CP
asymmetries of the decay modes $\Bpi$ and $\Bk$ are measured to be
\begin{eqnarray*} 
\begin{aligned}
\rpik&=\left\{
\begin{aligned}
&(3.90\pm0.06\pm0.04 ) \times10^{-2}
&\quad&{\rm for\,}7\tev\\
&(3.79\pm0.04\pm0.03 ) \times10^{-2}
&\quad&{\rm
for\,}8\tev\,,\\
\end{aligned}\right.\\
\apik&=\left\{
\begin{aligned}
& ( 1.92\pm1.53\pm0.15 ) \times10^{-2}
&\quad&{\rm for\,}7\tev\\
& ( 1.77\pm1.05\pm0.14 ) \times10^{-2}
&\quad&{\rm for\,}8\tev\,.\\
\end{aligned}\right.
\end{aligned}
\end{eqnarray*}
Here the first uncertainties are statistical, which are uncorrelated between the
7 and $8\tev$ results, and the second  uncertainties are
systematic, which are taken to be fully correlated between the 7 and
$8\tev$ results. The average of the 7 and 8\tev results, weighting each
according to its statistical uncertainty,
are
\begin{eqnarray*}
\begin{aligned}
\rpik&=(3.83\pm0.03\pm0.03 ) \times10^{-2}
,\\
\apik&=(1.82\pm0.86\pm0.14)\times 10^{-2}
. 
\end{aligned}
\end{eqnarray*}

The \lhcb collaboration has recently reported the \CP asymmetry
\mbox{$\ACP(\Bk)=(0.09\pm0.27\pm0.07)\times10^{-2}$} \cite{LHCb-PAPER-2016-054},
where the first uncertainty is statistical and the second systematic. The sample
analysed in Ref. \cite{LHCb-PAPER-2016-054} is statistically correlated with
that used in this analysis, but the correlation is only partial due to the use
of different trigger requirements. The correlation coefficient between the
statistical uncertainties of the two analyses is found to be $-4.8\%$. The
systematic uncertainty on $\ACP(\Bk)$ is taken to be uncorrelated with that on
the $\Delta\ACP$ measurement. Therefore the \CP asymmetry in the decay $\Bpi$ is
\begin{eqnarray*}
\ACP(\Bpi)=\apik+\ACP(\Bk)=(1.91\pm0.89\pm0.16)\times10^{-2}
. 
\end{eqnarray*}
This is the most precise determination of $\ACP(\Bpi)$ to date, and it
supersedes the previous \lhcb result \cite{LHCb-PAPER-2011-024}.  The $\rpik$
and $\ACP(\Bpi)$ measurements can be combined with measurements of decay rates
and \CP asymmetries in other $b\to c\bar{c}d$  decays, such as $B^0 \to \jpsi
\pi^0$, to understand the effect of loop contributions in  $b\to c\bar{c}s$
decays using SU(3) flavour symmetry\cite{Ligeti:2015yma,DeBruyn:2014oga}.

\section*{Acknowledgements}

\noindent We express our gratitude to our colleagues in the CERN
accelerator departments for the excellent performance of the LHC. We
thank the technical and administrative staff at the LHCb
institutes. We acknowledge support from CERN and from the national
agencies: CAPES, CNPq, FAPERJ and FINEP (Brazil); NSFC (China);
CNRS/IN2P3 (France); BMBF, DFG and MPG (Germany); INFN (Italy); 
FOM and NWO (The Netherlands); MNiSW and NCN (Poland); MEN/IFA (Romania); 
MinES and FASO (Russia); MinECo (Spain); SNSF and SER (Switzerland); 
NASU (Ukraine); STFC (United Kingdom); NSF (USA).
We acknowledge the computing resources that are provided by CERN, IN2P3 (France), KIT and DESY (Germany), INFN (Italy), SURF (The Netherlands), PIC (Spain), GridPP (United Kingdom), RRCKI and Yandex LLC (Russia), CSCS (Switzerland), IFIN-HH (Romania), CBPF (Brazil), PL-GRID (Poland) and OSC (USA). We are indebted to the communities behind the multiple open 
source software packages on which we depend.
Individual groups or members have received support from AvH Foundation (Germany),
EPLANET, Marie Sk\l{}odowska-Curie Actions and ERC (European Union), 
Conseil G\'{e}n\'{e}ral de Haute-Savoie, Labex ENIGMASS and OCEVU, 
R\'{e}gion Auvergne (France), RFBR and Yandex LLC (Russia), GVA, XuntaGal and GENCAT (Spain), Herchel Smith Fund, The Royal Society, Royal Commission for the Exhibition of 1851 and the Leverhulme Trust (United Kingdom).

\addcontentsline{toc}{section}{References}
\setboolean{inbibliography}{true}
\bibliographystyle{LHCb}
\bibliography{main,LHCb-PAPER,LHCb-CONF,LHCb-DP,LHCb-TDR,myPaper}

\newpage

\newpage
\centerline{\large\bf LHCb collaboration}
\begin{flushleft}
\small
R.~Aaij$^{40}$,
B.~Adeva$^{39}$,
M.~Adinolfi$^{48}$,
Z.~Ajaltouni$^{5}$,
S.~Akar$^{59}$,
J.~Albrecht$^{10}$,
F.~Alessio$^{40}$,
M.~Alexander$^{53}$,
S.~Ali$^{43}$,
G.~Alkhazov$^{31}$
% ...

{\footnotesize \it
% ...
$ ^{5}$Clermont Universit{\'e}, Universit{\'e} Blaise Pascal, CNRS/IN2P3, LPC, Clermont-Ferrand, France\\
% ...
$ ^{10}$Fakult{\"a}t Physik, Technische Universit{\"a}t Dortmund, Dortmund, Germany\\
% ...
$ ^{31}$Petersburg Nuclear Physics Institute (PNPI), Gatchina, Russia\\
% ...
$ ^{39}$Universidad de Santiago de Compostela, Santiago de Compostela, Spain\\
$ ^{40}$European Organization for Nuclear Research (CERN), Geneva, Switzerland\\
% ...
$ ^{43}$Nikhef National Institute for Subatomic Physics, Amsterdam, The Netherlands\\
% ...
$ ^{48}$H.H. Wills Physics Laboratory, University of Bristol, Bristol, United Kingdom\\
% ...
$ ^{53}$School of Physics and Astronomy, University of Glasgow, Glasgow, United Kingdom\\
% ...
$ ^{59}$University of Cincinnati, Cincinnati, OH, United States\\
% ...
}
\end{flushleft}

\end{document}